\documentclass[useAMS,usenatbib,usegraphicx]{mn2e}

\usepackage{url}
\usepackage[varg]{txfonts}
\usepackage{color}

\renewcommand\vec[1]{\bmath{#1}}

\newcommand{\veceta}{\vec{\eta}}
\newcommand{\vecalp}{\vec{\alpha}}

\newcommand{\zl}{z_{\mathrm{l}}}
\newcommand{\zs}{z_{\mathrm{s}}}

\newcommand{\cauldron}{\textsc{cauldron}}
\newcommand{\fvfps}{\textsc{fvfps}} 
\newcommand{\galaxy}{SDSS\,J2321$-$097}
\newcommand{\Lz}{L_{z}}

\newcommand{\mt}{\tilde{m}}
\newcommand{\talp}{\alpha_{0}}
\newcommand{\PA}{\vartheta_{\mathrm{PA}}}

\newcommand{\slope}{\gamma'}
\newcommand{\shear}{\zeta}
\newcommand{\angshear}{\vartheta_{\zeta}}

\newcommand{\Reff}{R_{\mathrm{e}}}

\newcommand{\REin}{R_{\mathrm{Einst}}}

\newcommand{\xy}{$xy$-plane}
\newcommand{\yz}{$yz$-plane}
\newcommand{\zx}{$zx$-plane}
\newcommand{\Rcore}{R_{\mathrm{s}}}

\title{Crash-testing the CAULDRON code for joint lensing and dynamics analysis of early-type galaxies}

\author[M. Barnab\`e et al.]{%
  Matteo Barnab\`e$^{1}$\thanks{E-mail: M.Barnabe@astro.rug.nl}, 
  Carlo Nipoti$^{2}$,
  L\'eon V. E. Koopmans$^{1}$,
  Simona Vegetti$^{1}$
  \newauthor
  and Luca Ciotti$^{2}$\\
  $^{1}$Kapteyn Astronomical Institute, University of Groningen, 
  PO Box 800, 9700\,AV Groningen, the Netherlands\\
  $^{2}$Astronomy Department, University of Bologna, via Ranzani 1,
  40127 Bologna, Italy}

\begin{document}

\date{Published as MNRAS 393, 1114}

\maketitle

\label{firstpage}

\begin{abstract}
  We apply the joint lensing and dynamics code for the analysis of
  early-type galaxies, {``\cauldron''}, to a rotating N-body stellar
  system with dark matter halo which significantly violates the two
  major assumptions of the method, i.e. axial symmetry supported by a
  two-integral distribution function. The goal is to study how
  {\cauldron} performs in an extreme case, and to determine which
  galaxy properties can still be robustly recovered. Three data sets,
  corresponding to orthogonal lines of sight, are generated from the
  N-body system and analysed with the identical procedure followed in
  the study of real lens galaxies, adopting an axisymmetric power-law
  total density distribution. We find that several global properties
  of the N-body system are recovered with remarkable accuracy, despite
  the fact that the adopted power-law model is too simple to account
  for the lack of symmetry of the true density distribution. In
  particular, the logarithmic slope of the total density distribution
  is robustly recovered to within less than $10\%$ (with the exception
  of the ill-constrained very inner regions), the inferred
  angle-averaged radial profile of the total mass closely follows the
  true distribution, and the dark matter fraction of the system
  (inside the effective radius) is correctly determined within $\sim
  10\%$ of the total mass. Unless the line of sight direction is
  almost parallel to the total angular momentum vector of the system,
  reliably recovered quantities also include the angular momentum, the
  $V/\sigma$ ratio, and the anisotropy parameter $\delta$. We conclude
  that the {\cauldron} code can be safely and effectively applied to
  real early-type lens galaxies, providing reliable information also
  for systems that depart significantly from the method's assumptions.
\end{abstract}

\begin{keywords}
  gravitational lensing ---
  methods: N-body simulations ---
  galaxies: elliptical and lenticular, cD --- 
  galaxies: kinematics and dynamics --- 
  galaxies: structure.
\end{keywords}


\section{Introduction}
\label{sec:introduction}

Determining the structure of early-type galaxies and reliably pinning
down their dark matter content is a crucial step in order to fully
understand the formation and evolution processes of these
systems. 

Within the currently favoured cosmological scenario, the $\Lambda$CDM
paradigm, early-type galaxies are thought to be formed via
hierarchical merging of lower mass galaxies \citep{Toomre1977,
White-Frenk1991, Barnes1992, Cole2000}. While very successful in
reproducing many observational features of elliptical galaxies,
including complex ones \citep[see e.g.][]{Jesseit2007}, these
formation models are still encountering difficulties in explaining the
origin of the empirical scaling laws that correlate the global
properties of early-type galaxies \citep[see
e.g.][]{Robertson2006}. Providing a reliable and detailed description
of the mass density distribution, orbital structure and intrinsic
properties of early-type galaxies is therefore critical in order to
enable stringent tests of galaxy formation models.

For this reason, considerable effort has been devoted during the last
decades towards the observation and the modelization of nearby
early-type galaxies, by means of both stellar dynamics and X-ray
studies, finding more or less strong evidence for a dark matter halo
component (e.g. \citealt{Fabbiano1989}, \citealt{Mould1990},
\citealt*{Saglia1992}, \citealt{Bertin1994}, \citealt{Franx1994},
\citealt{Carollo1995}, \citealt{Arnaboldi1996}, \citealt{Rix1997},
\citealt{Matsushita1998}, \citealt{Loewenstein1999},
\citealt{Gerhard2001}, \citealt{Borriello2003},
\citealt{Romanowsky2003}, \citealt{Humphrey2006}, \citealt{Forbes2008}
and more recently the SAURON collaboration: see
e.g. \citealt{SauronII}, \citealt{SauronIII},
\citealt{SauronIV}). Both methods, however, present some difficulties.
In the case of stellar dynamics it is believed that some degeneracy
can be present between the mass profile of the galaxy and the
anisotropy of the stellar velocity dispersion tensor, which can be
alleviated when higher order velocity moments are available \citep[see
e.g.][]{Gerhard1993} or by using physically motivated distribution
functions \citep[see e.g.][for a discussion of this
point]{Bertin2000}. X-ray analyses, on the other hand, can seriously
overestimate the total mass of the system if the assumption of
hydrostatic equilibrium for the hot gas does not hold, especially near
the center \citep[see e.g.][]{Pellegrini2006, Ciotti-Pellegrini2008}.

A full understanding of the evolution of early-type galaxies cannot be
achieved without extending the study also to the mass density profile
of objects at higher redshift ($z \ga 0.1$). This, however, has not
been attempted until recently, due to observational limitations and to
the increased difficulty in extracting detailed kinematic information,
which hinders traditional analyses based on stellar dynamics only. An
effective solution in order to overcome these issues is constituted by
a joint analysis which combines the constraints from stellar dynamics
with the information obtained from gravitational lensing, when the
early-type galaxy also happens to act as a lens with respect to a
background source at higher redshift \citep{Treu-Koopmans2002,
Koopmans-Treu2002, Treu-Koopmans2003, Treu-Koopmans2004,
vandeVen2008}.  \citet{Koopmans2006} have successfully used this
combined approach to analyse fifteen early-type lens galaxies (within
a redshift range $z = 0.06 - 0.33$) discovered in the Sloan Lens ACS
Survey (SLACS, \citealt{Bolton2006}) as well as six systems between $z
\sim 0.5$ and $1$ from the Lenses Structure and Dynamics (LSD) Survey,
showing that all of the examined systems are well described by a
power-law total density distribution very close to $r^{-2}$. The
technique for the joint lensing and dynamics analysis has been
expanded by \citet[][hereafter BK07]{Barnabe-Koopmans2007} into a
general and self-consistent method, completely embedded within the
framework of Bayesian statistics, which puts constraints on the total
density distribution of the lens galaxy by taking advantage of all the
available data, i.e. not only the lensed image and a single stellar
velocity dispersion measurement, but also the surface brightness
distribution and the 2D kinematic maps (first and second projected
velocity moments).

Similar to other methods for the determination of the structure and
internal dynamics of early-type galaxies, already mentioned above, the
joint lensing and dynamics analysis also relies on a certain number of
assumptions. For example, the simple and robust approach of
\citet{Koopmans2006} treats the gravitational lensing and the stellar
dynamics as independent problems. The projected mass distribution of
the lens galaxy is modelled as a singular isothermal ellipsoid in order
to determine the total mass within the Einstein radius, which is then
used as a constraint for the dynamical model, where spherical symmetry
and a specific prescription for the stellar orbital anisotropy are
assumed. The more sophisticated framework of BK07 is designed to be
very general and allows in principle for an arbitrary choice of the
total potential. In practice, however, such freedom must balance
against technical and computational limitations. Therefore, in order
to have a fast and efficient algorithm, the current implementation of
the method, the {\cauldron} code\footnote{Combined Algorithm for
Unified Lensing and Dynamics \mbox{ReconstructiON}}, is restricted to
axisymmetric potentials and two-integral stellar phase-space
distribution functions (DFs). Under these hypotheses, it has been
shown in BK07 that the method is capable of recovering with
considerable accuracy the correct potential parameters and inclination
angle, even in the presence of realistic noise.

The point above raises the question of whether (and to what extent)
the simplifying assumptions can be deemed valid for the astrophysical
systems to which such methods are applied.  In fact, real galaxies are
not idealized objects, and there is no reason to expect them to be
exactly axisymmetric (and neither triaxial ellipsoids) or to have two
or three integrals of motion. Whereas axial symmetry generally seems
to constitute a fairly good approximate description for most
early-type galaxies, a more detailed inspection, such as that allowed
by the SAURON observations \citep[see][]{SauronIII, SauronVIII},
reveals a multitude of features indicating departure from axisymmetry,
e.g. the presence of isophotal twist in the surface brightness
distribution, minor axis rotation and kinematically decoupled cores.

For this reason, in the present paper we apply our algorithm to the
end-product of a two-component (stars plus dark matter) N-body
simulation of a merger process, i.e. to a system which does not obey
any restrictive prescription of symmetry, and therefore violates the
assumptions of the method. We aim to study how the {\cauldron}
algorithm performs when subjected to this kind of ``crash-test'', and
which quantities can be robustly recovered even in such an extreme
case. A similar approach has been followed by \citet{Thomas2007}, who
have applied their three-integral axisymmetric Schwarzschild code to
the study of non axisymmetric N-body merger remnants, although without
any gravitational lensing information, and by \citet{Meneghetti2007}
in the case of clusters of galaxies.

The paper is organized as follows. In Section~\ref{sec:code} we
provide an overview of the {\cauldron} algorithm for combined lensing
and dynamics analysis. In Section~\ref{sec:simulation} we summarize
the properties of the N-body system that we use as lens galaxy. In
Section~\ref{sec:observables} we describe how the 2D maps of the
simulated data with added realistic noise are obtained from the
particle distribution. In Section~\ref{sec:results} we apply the joint
lensing and dynamics analysis to the simulated data and we present the
results, which are then further discussed in
Section~\ref{sec:conclusions}, where also conclusions are drawn.


\section{The cauldron algorithm for joint lensing and dynamics analysis}
\label{sec:code}

In this Section we recall the main features of the {\cauldron}
algorithm. We refer the reader to BK07 for a fully detailed
description of the method.

The central tenet of a self-consistent joint analysis is to adopt a
total gravitational potential $\Phi$ (or, equivalently, the total
density profile $\rho$, from which $\Phi$ is calculated via the
Poisson equation), parametrized by a set $\veceta$ of non-linear
parameters, and use it simultaneously for both the gravitational
lensing and the stellar dynamics modelling of the data. As shown in
BK07, these two modelling problems, while different from a physical
point of view, can be expressed in an analogous way as a single set of
coupled (regularized) linear equations. For any given choice of the
non-linear parameters, the equations can be solved (in a direct,
non-iterative way) to simultaneously obtain as the best solution for
the chosen potential model: (i) the unlensed source surface brightness
distributions, and (ii) the weights of the elementary stellar dynamics
building blocks \citep[e.g.\ orbits or two-integral components,
TICs,][]{Schwarzschild1979, Verolme-deZeeuw2002}.  This linear
optimization scheme is consistently embedded in the framework of
Bayesian statistics. As a consequence, it is possible to objectively
assess the probability of each model by means of the evidence merit
function and, therefore, to rank different models
\citep[see][]{MacKay1992, MacKay1999, MacKay2003}. In this way, by
maximizing the evidence, it is possible to recover the set of
non-linear parameters $\veceta$ corresponding to the ``best''
potential model.  Here, in the context of Bayesian statistics, the
``best model'' means the most plausible model in an Occam's razor
sense, given the data and the adopted form of the regularization (the
optimal level of the regularization is also set by the evidence).

Whereas the method is in principle extremely general\footnote{One
could adopt for example a completely general pixelized potential for
which the best profile is then determined via Bayesian statistics only
by the data.}, its current practical implementation, which will be
referred to as the {\cauldron} algorithm, is more restricted in order
to make it computationally efficient and applies specifically to
axisymmetric potentials, $\Phi(R,z)$, and two-integral DFs $f = f(E,
\Lz)$ (where $E$ and $\Lz$ are, respectively, the energy and the
angular momentum along the rotation axis). Under these assumptions,
the dynamical model can be constructed by making use of the fast BK07
numerical implementation of the two-integral Schwarzschild method
developed by \citet{Cretton1999} and \citet{Verolme-deZeeuw2002},
whose building blocks are not stellar orbits (as in the classical
Schwarzschild method) but TICs.\footnote{A TIC can be visualized as an
elementary toroidal system, completely specified by a particular
choice of energy $E$ and axial component of the angular momentum
$\Lz$. TICs have simple $1/R$ radial density distributions and
analytic unprojected velocity moments, and by superposing them one can
build $f(E, \Lz)$ models for arbitrary spheroidal potentials
\citep[cf.][]{Cretton1999}: all these characteristics contribute to
make TICs particularly valuable and ``inexpensive'' building blocks
when compared to orbits.}  The weights map of the optimal TIC
superposition which best reproduces the observables is yielded as an
outcome of the joint analysis.

The {\cauldron} algorithm has been successfully tested against the
analytic power-law galaxy models of \citet{Evans1994}, which respect
by construction the assumptions of axisymmetry and two-integral DF,
and afterwards has been employed for the detailed analysis of the
SLACS lens galaxy {\galaxy} \citep[][hereafter C08]{Czoske2008}. The
latter is a case study which presents a benchmark data set
particularly well suited to the needs of {\cauldron},
i.e. high-resolution HST/ACS images of the gravitational lensed source
and of the surface brightness distribution of the lens galaxy, and 2D
maps of the projected velocity moments of the lens galaxy derived from
VLT-VIMOS observations. Therefore, the observations of {\galaxy} will
be used as a reference in order to generate the simulated observables
for the present study (see Section~\ref{sec:observables}).

It has been shown by the work of \citet{Koopmans2006} that a simple
power-law model for the total density distribution provides a
satisfactory description for all of the SLACS lens galaxies examined
so far. This has been further confirmed in the case of {\galaxy},
where a fully self-consistent analysis was performed. In the present
work, we aim to study the simulated galaxies exactly as we would do
for real objects, without assuming any \mbox{a priori} knowledge, and
therefore we still adopt the same power-law model that has been used
in C08. In particular, the total mass density distribution of the
galaxy is taken to be a power-law stratified on axisymmetric
homoeoids:
\begin{equation}
  \label{eq:rho}
  \rho(m) = \frac{\rho_{0}}{m^{\slope}}, \quad 0 < \slope < 3,
\end{equation}
where $\rho_{0}$ is a density scale, $\slope$ will be referred to as
the (logarithmic) slope of the density profile, and
\begin{equation}
  \label{eq:m}
  m^2 = \frac{R^2}{a_0^2} + \frac{z^2}{c_0^2} 
  = \frac{R^2}{a_0^2} + \frac{z^2}{a_0^2 q^2} ,
\end{equation}
where $c_0$ and $a_0$ are length-scales and $q\equiv c_0/a_0$.

The (inner) gravitational potential associated with a homoeoidal
density distribution $\rho(m)$ is given by \citet{Chandrasekhar1969}
formula. In our case, for $\slope \ne 2$, one has
\begin{equation}
  \label{eq:pot}
  \Phi(R,z) = - \frac{\Phi_{0}}{\slope-2} \int_{0}^{\infty} 
  \frac{{\mt}^{2-\slope}}
  {(1+\tau) \sqrt{q^2 + \tau}}  \,\mathrm{d} \tau\;,
\end{equation}
while for $\slope = 2$ 
\begin{equation}
  \label{eq:pot.2}
  \Phi(R,z) = \Phi_{0} \int_{0}^{\infty} 
  \frac{\log \mt}
  {(1+\tau) \sqrt{q^2 + \tau}}\,  \mathrm{d} \tau ,
\end{equation}
where $\Phi_{0} = 2 \pi G q a_{0}^2 \rho_{0}$ and
\begin{equation}
  \label{eq:mt}
  \mt^{2} = \frac{R^2}{a_{0}^2 (1+\tau)} + \frac{z^2}{a_{0}^2 (q^2+\tau)} .
\end{equation}

There are three free non-linear parameters in the potential to be
determined via the evidence maximization: $\Phi_{0}$ (or equivalently,
through equation [B4] of BK07, the lens strength $\talp$), the slope
$\slope$ and the axial ratio $q$. When required by the data, it is
straightforward to include a core radius $\Rcore$ in the
density distribution. Beyond the previously mentioned parameters,
there are four additional parameters which determine the geometry of
the observed system: the position angle $\PA$, the inclination $i$ and
the coordinates of the centre of the lens galaxy with respect to the
sky grid. The position angle and the lens center can usually be
accurately determined by means of a preliminary exploration and kept
fixed afterwards in order to reduce the number of free
parameters. Finally, for a proper modelling of the lensed image it can
be necessary to include two more parameters (shear strength $\shear$
and shear angle $\angshear$) in order to account for external shear.

A curvature regularization (as described in \citealt{Suyu2006} and
appendix~A of BK07) is adopted for both the gravitational lensing and
the stellar dynamics reconstructions. As discussed in BK07, the
initial guess values of the hyperparameters (defining the level of the
regularization) are chosen to be quite large, since the convergence to
the maximum is faster when starting from an overregularized system.


\section{The N-body system}
\label{sec:simulation}

\begin{figure}
  \centering
  \resizebox{1.00\hsize}{!}{\includegraphics{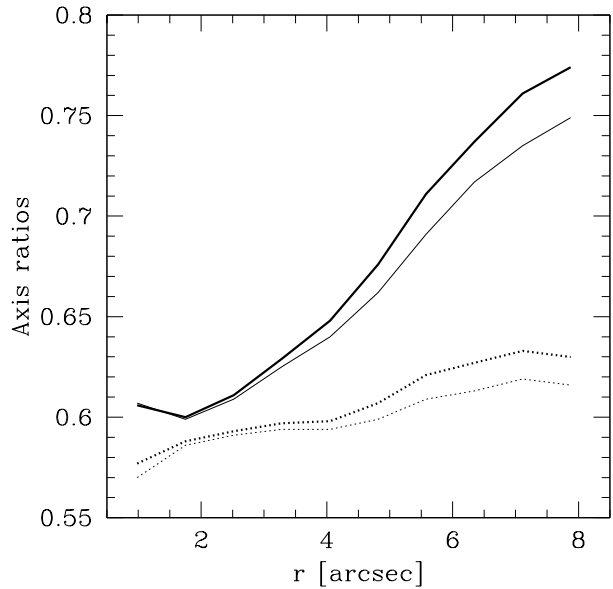}}
  \caption{Axis ratios $b/a$ (solid curves) and $c/a$ (dotted curves)
    as functions of radius for the stellar (thin curves) and total
    (thick curves) density distributions of the N-body system used as
    lens (we assumed $r_*=1.969$ arcsec).}
  \label{fig:axisratio}
\end{figure}

The model galaxy that we use as lens in the present work is the
end-product of a numerical N-body simulation of a dissipationless
merging between two equal-mass spherical galaxies embedded in their
dark matter halos.  The simulation, run with the treecode {\fvfps}
(Fortran Version of a Fast Poisson Solver; \citealt*{Londrillo2003};
\citealt*{Nipoti2003}), has been already presented and described in a
previous paper \citep*{Nipoti2007}, in which it is named E25o. Here we
recall the main properties of the end-product of this simulation,
while we refer the reader to \citet{Nipoti2007} for a detailed
description of the initial conditions.

The simulation end-product is a nearly ellipsoidal virialized system,
comprising a stellar component and a dark-matter component.  The total
number of particles is $\sim 1.2 \times 10^6$, and all particles have
the same mass. The stellar component has total mass $\sim 2 \, M_*$
and angle-averaged half-mass radius $\sim 3.8 \, r_*$, where $M_*$ and
$r_*$ are the code length and mass unit. The dark matter component is
more massive and more extended than the stellar component, having mass
$\sim 10 \, M_*$ and half mass radius $\sim 19.3 \, r_*$. The system
has a virial velocity dispersion $\sim 0.55 (G M_*/r_*)^{1/2}$ and
non-vanishing total angular momentum ${\bf L}$, corresponding to a
value $\lambda \sim 0.07$ in terms of the spin parameter $\lambda
\equiv {|E_{\rm tot}|^{1/2} ||{\bf L}|| G^{-1} M_{\rm tot}^{-5/2}}$,
where $E_{\rm tot}$ is the total energy, and $M_{\rm tot}\sim 12 \,
M_*$ is the total mass. Of course, the model can be rescaled to
represent physical systems of any size and mass by choosing proper
values of $M_*$ and $r_*$.

We define the minor-to-major ($c/a$) and intermediate-to-major ($b/a$)
axis ratios of the system at a radius $r$ as the corresponding axis
ratios of the inertia ellipsoid of particles within an ellipsoid of
angle-averaged radius $r$ \citep*[for details, see][]{Nipoti2002}.  
We find that the system is triaxial, with direction of the principal
axes and axis ratios depending on radius. Figure~\ref{fig:axisratio}
shows $b/a$ (solid curve) and $c/a$ (dotted curve) as functions of
radius for the stellar (red) and total (black) density distributions
within about the half-mass radius of the stellar component (we assumed
$r_*=1.969$ arcsec, see Section~\ref{sec:observables}). Both the
stellar and the total distribution are strongly triaxial at $r \ga 5$
arcsec and mildly triaxial (almost prolate) at $r \la 5$ arcsec. The
angle-averaged total density and mass profiles of the model are
plotted as solid black curves in Figs.~\ref{fig:prof_dens}
and~\ref{fig:prof_mass}, respectively.


\section{Construction of the observables}
\label{sec:observables}

In this Section we detail how the simulated observables (or ``mock
data'') are generated from the dark matter and stellar particle
distribution taken from the virialized end-product of the N-body
simulation described in Section~\ref{sec:simulation}.

In order to convert the N-body simulation in a realistic data set with
plausible physical characteristics (effective radius, redshift of lens
galaxy and source, Einstein radius), we use as a reference the actual
lens galaxy {\galaxy}, for which both data analysis and joint lensing
and dynamics study are presented in C08. Therefore, we will adopt for
the noise level and the sampling of the data (i.e. size and binning of
the data grids) the corresponding values of {\galaxy}.

\begin{figure}
  \centering
  \resizebox{1.00\hsize}{!}{\includegraphics[angle=-90]{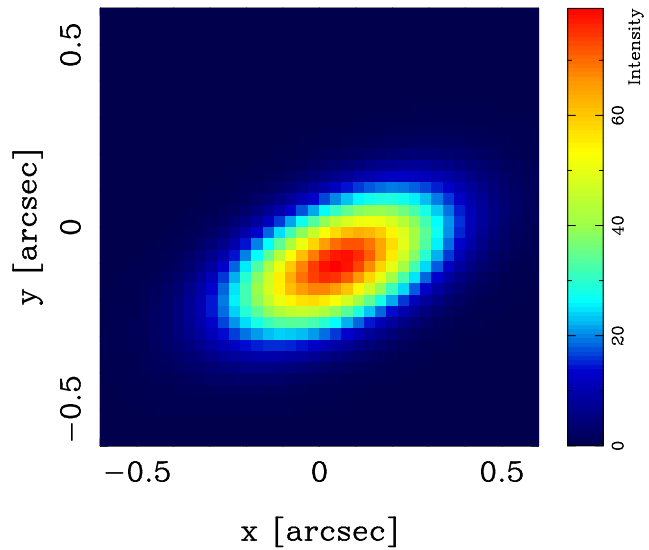}}
  \caption{Mock source to be lensed. This is an elliptical Gaussian
  brightness distribution with $\sigma_{x} = 0.2$ arcsec, $\sigma_{y}
  = 0.1$ arcsec, position angle $\PA = 115\degr$ and with a slight
  offset with respect to the lens center.}
  \label{fig:LENsource}
\end{figure}

First, we impose for the simulated galaxy the same redshift of
{\galaxy}, i.e. $\zl = 0.0819$. The adopted source
(Fig.~\ref{fig:LENsource}) to produce the artificial lensed image is
an elliptical Gaussian distribution, slightly offset with respect to
the center of mass of the lens galaxy and (again in analogy with
{\galaxy}) located at a redshift $\zs = 0.5342$.  We then fix the
values $r_* = 3$ kpc ($\simeq 1.969$ arcsec at redshift $\zl$) and
$M_* = 3 \times 10^{11} \, M_{\odot}$ for, respectively, the length
and mass units of the simulated galaxy.  This setup produces a galaxy
that, when is observed along an arbitrary line of sight, displays
realistic values for both the effective radius $\Reff$ ($\sim 5 - 6$
arcsec, corresponding to $\sim 8$ kpc at the redshift~$\zl$), and for
the Einstein radius $\REin$ ($\sim 2$ arcsec).

With this choice of $r_{*}$ and $M_{*}$, the simulated galaxy is a
massive system of total mass $3.6 \times 10^{12} \, M_{\sun}$, with a
dark halo which extends up to several hundred kpc. In analogy with the
data set of {\galaxy}, the spatial coverage of our data is confined to
the very inner regions of the galaxy, approximately corresponding to
$\Reff/2$. However, a fair amount of information comes also from more
distant regions of the system which are seen in projection along the
line-of-sight (see discussion in C08).

We define for the galaxy an orthogonal reference frame with the origin
in the center of mass of the simulated system; the $z$ axis is
oriented along the direction of the total angular momentum of the
stellar component. Since all the observables are quantities projected
in the sky plane, we select three orthogonal lines of sight along which
the simulated galaxy is assumed to be observed. In this way, from a
single simulation, we obtain three different data sets. The first line
of sight is chosen to be approximately oriented along the $z$ axis of
the galaxy, and therefore in the following we will refer to the
corresponding projection as the {\xy} projection or, for simplicity,
the ``face-on'' projection. The remaining two projections are taken
along mutually orthogonal axes both perpendicular to the first line of
sight and will be called the $yz$- and {\zx} projections or, for
brevity sake, the ``edge-on'' projections.

The circularized half-light radii for the three projections are found
to be $\Reff = 5.227$ arcsec ({\yz}), $5.359$ arcsec ({\zx}) and
$6.156$ arcsec ({\xy}).

When considering any one of the three projections, we will always
indicate as $z'$ the direction of the particular line-of-sight. For
each projection, the observables are then calculated following the
same procedure:

\begin{enumerate}

\item \emph{Lensed image.} All the particles (both stellar and dark
matter ones), properly weighted by the respective masses, are cast in
projection along the line-of-sight and binned in a 2D-grid along the
sky plane in order to generate from the total density
$\rho_{\mathrm{tot}}$ a projected density map
$\Sigma_{\mathrm{tot}}$. The latter, once normalized to the critical
density $\Sigma_{\mathrm{cr}} = (c^{2}/4 \pi G) \, (D_{\mathrm{s}}/
D_{\mathrm{d}} D_{\mathrm{ds}})$, a quantity which depends only on
geometry of the system\footnote{$D_{\mathrm{s}}$, $D_{\mathrm{l}}$ and
$D_{\mathrm{ls}}$ are the angular diameter distances from the observer
to the source, from the observer to the lens and from the lens to the
source, respectively.}, yields the convergence field $\kappa$ on the
projection plane. The two-dimensional Poisson equation $\nabla^{2}
\psi = 2 \kappa$ relates the convergence to the projected potential
$\psi$, whose gradient immediately gives the deflection angle vector
field: $\vecalp = \nabla \psi$ \citep*[see e.g.][]{SEF1992}.

We make use of a square grid of $3000 \times 3000$ bins for the
convergence.  The grid size (equivalent to $\sim 40$ arcsec) is such
that it contains, in projection, about half of the total number of
particles. Including more distant particles does not have any
discernible effect on the outcome, except slowing down all the related
calculations, since the number of bins must be increased in order to
keep the resolution constant. The Poisson equation is then solved via
fast Fourier transform, using the freely available package FFTW
\citep{FFTW}, and enabling us to obtain, from the convergence
(appropriately padded in order to avoid numerical issues with the
Fourier transform), the two components of deflection angle
$\vecalp$. With these deflection angle maps and the mock source
described above, the {\cauldron} code can straightforwardly generate
the lensed image, already convolved with the HST/ACS F814W point
spread function (PSF) obtained with \textsc{tiny tim}
\citep{Krist1993}. The lensed image is constructed (in analogy with
the lensing data for {\galaxy}) on $100 \times 100$ grid, with each
pixel corresponding to $0.05$ arcsec. Finally, a noise distribution
based on the covariance maps of the HST images of {\galaxy} is added
to the lensed image map in order to produce the final data set.

The obtained lensing data sets for the three projections are shown in
the upper right panel of Figs~\ref{fig:YZ-LEN}, \ref{fig:ZX-LEN}
and~\ref{fig:XY-LEN}.

\item \emph{Surface brightness distribution.} We assume
that the stellar mass-to-light ratio is independent of position. The
stellar particles are cast along the chosen line-of-sight on a 2D-grid
in the sky plane. Such grid is padded and oversampled by a factor of
$3$ with respect to the final grid adopted for this observable (see
later). This is necessary since, in order to take into account the
effect of seeing, we have to convolve this quantity with the PSF (here
modelized as a Gaussian distribution of FWHM $= 0.10$ arcsec). The map
is then resampled, generating the surface brightness distribution of
the galaxy on a $50 \times 50$ grid ($1$ pixel $ = 0.10$ arcsec). Mock
noise is then added consistently with the corresponding variance maps
of {\galaxy}. These images are to first-order equivalent to HST-NICMOS
images.

The obtained data sets for the surface brightness distribution of the
three projections are shown in the upper left panel of
Figs~\ref{fig:YZ-DYN}, \ref{fig:ZX-DYN} and~\ref{fig:XY-DYN}.

\item \emph{Line-of-sight projected velocity moments.}  
The procedure to generate the kinematic maps is similar to the one
described above for the surface brightness distribution, including the
oversampling and convolution with the Gaussian PSF (but with a broader
FWHM $= 0.90$ arcsec, typical for ground-based observations with the
VLT).

The only difference is that now the velocities projected along the
line-of-sight ($v_{z'}$ and $v^{2}_{z'}$ for the two kinematic maps
respectively) associated with each particle are summed up on each
cell, producing the unweighted maps to be used by {\cauldron}. These
maps can be divided by the surface brightness distribution sampled on
the same grid (the ``kinematic'' grid) in order to obtain the weighted
maps, i.e. the quantities $\langle v_{z'} \rangle$ and $\langle
v^{2}_{z'} \rangle$. The projected velocity dispersion is obtained as
$\sigma^{2}_{z'} = \langle v^{2}_{z'} \rangle - {\langle v_{z'}
\rangle}^{2}$.

Due to the challenges of spectroscopic observations of distant
early-type galaxies, the kinematic grids of {\galaxy} have only $9
\times 9$ elements, with $1$ pixel $ = 0.67$ arcsec (i.e. the
VLT-VIMOS IFU fiber size). Since we are mimicking those observation,
we adopt the same grid.  As usual, the mock noise is added according
to what we know from {\galaxy}.

The obtained kinematic data sets for the three projections are
presented in the upper central (for the line-of-sight velocity) and
upper right (for the line-of-sight velocity dispersion) panels of
Figs~\ref{fig:YZ-DYN}, \ref{fig:ZX-DYN} and~\ref{fig:XY-DYN}.

\end{enumerate}



\section{Analysis and results}
\label{sec:results}

In this section we illustrate the results of the joint lensing and
dynamics analysis performed with {\cauldron} on the three data sets
generated (as described in Section~\ref{sec:observables}) from
orthogonal projections of the simulated galaxy.

\begin{table}
  \centering
  \caption{Recovered non-linear parameters of the best power-law
    models for the three data sets (obtained as projection of the
    simulated systems on the three orthogonal planes indicated in the
    columns): inclination $i$ (in degrees), shear strength $\shear$
    and angle $\angshear$ (in degrees), lens strength $\talp$,
    logarithmic slope $\slope$ and flattening $q$.}
  \smallskip
  \begin{tabular}{ c c c c }
    \hline
    \noalign{\smallskip}
     & {\yz} & {\zx} & {\xy} \\
    \noalign{\smallskip}
    \hline
    \noalign{\smallskip}
    $i$         & 56.5                    & 88.9                    & 53.6 \\
    $\shear$    & $ 5.36 \times 10^{-2} $ & $ 8.71 \times 10^{-2} $ & $ 5.99 \times 10^{-2} $ \\
    $\angshear$ & $ -24.6 $               & 32.4                    & 27.8 \\
    $\talp$     & 0.652                   & 0.642                   & 0.520 \\
    $\slope$    & 2.214                   & 2.078                   & 2.234 \\
    $q$         & $ 1.000 $               & $ 1.124 $               & 0.864 \\
    \noalign{\smallskip}
    \hline
  \end{tabular}
  \label{tab:eta}
\end{table}


\subsection{Recovered structure}
\label{ssec:structure}

As discussed in Section~\ref{sec:code}, we have adopted an
axisymmetric power-law profile as a model for the total density
distribution. The recovered non-linear parameters of the best models
for the three data sets are reported in Table~\ref{tab:eta}. The
non-linear parameters listed in the Table are the inclination $i$ (in
degrees), the shear strength $\shear$ and angle $\angshear$ (in
degrees), the lens strength $\talp$, the logarithmic slope $\slope$
and the flattening $q$ (a $q$ larger than $1$ denotes a prolate
axisymmetric shape).  In order to speed up the optimization routine,
the best values for the lens center and galaxy position angle were
determined via preliminary runs, and afterwards were kept fixed (this
procedure is commonly used in this kind of analysis: see
e.g. C08). The core radius $\Rcore$ was initially set free as an
additional non-linear parameter, but, in order to make the
optimization faster, it was later fixed to a negligibly small value
after verifying that the introduction of this new parameter did not
lead to an increase of the value of the evidence merit function (and
therefore in a Bayesian sense the additional complexity was not
justified). The optimization routine yields also the best value for
the three hyperparameters which set the ideal level of regularization.

The reconstructed observables for gravitational lensing and stellar
dynamics that correspond to the best model for the {\yz} data set are
presented and compared to the data in Figs~\ref{fig:YZ-LEN}
and~\ref{fig:YZ-DYN}, respectively. Analogously,
Figs.~\ref{fig:ZX-LEN}-\ref{fig:ZX-DYN}
and~\ref{fig:XY-LEN}-\ref{fig:XY-DYN} show the same quantities for the
{\zx} and {\xy} (i.e. face-on projection) data sets.

It is apparent that the residuals of the reconstructed lensed image
are fairly large, in particular in the cases of the $zx$- and {\xy}
projections, where moreover the reconstructed source appears patchy
and unrealistically pixelized, despite the regularization. The same
effects, while less pronounced, are discernible also in the case of
the {\yz} projection, where the reconstruction was most
successful. This is a clear indication that the underlying total
density distribution is actually more complex or inhomogeneous than
the simple power-law profile that we are using as a model (see
Section~\ref{ssec:source} for a more extended discussion) and that the
model is unable to de-lens all lensed images simultaneously into a
single well-defined source.

\begin{figure}
  \centering
  \resizebox{1.00\hsize}{!}{\includegraphics[angle=-90]{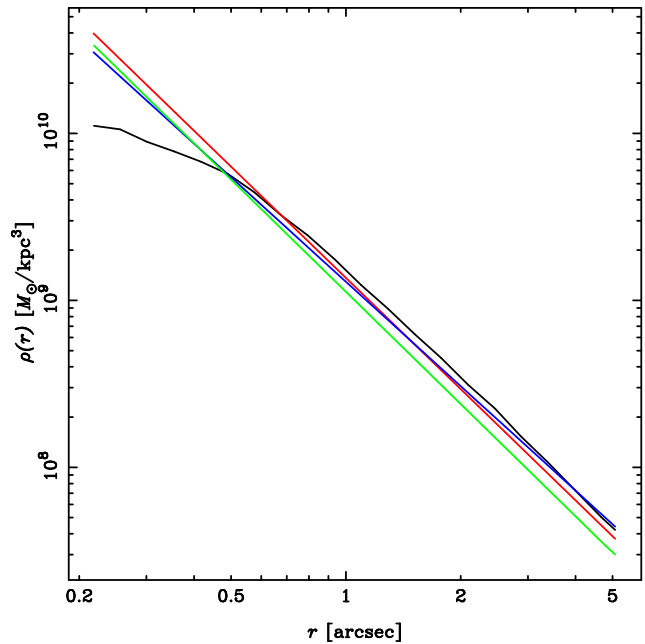}}
  \caption{Angle-averaged total density profile of the simulated system
  (black line) compared with the density profiles of the recovered
  best models (see Table~\ref{tab:eta}) for the three orthogonal
  projections data sets. The {\yz} model (red line) has a
  logarithmic density slope $\slope_{yz} = 2.214$, the {\zx}
  model (blue line) has a slope $\slope_{zx} = 2.078$, and the
  {\xy} model (green line) has a slope $\slope_{xy} = 2.234$.}
  \label{fig:prof_dens}
\end{figure}

The surface brightness distribution and the kinematics appear to be
reasonably well reconstructed. However, in the inner region the
reconstructed velocity dispersion is more peaked than in the data,
with the possible exception of the {\zx} data set (the one for which
the recovered slope is the most shallow,
cf. Table~\ref{tab:eta}). Together with the faint central image
visible in all the lensing data sets, this indicates that the total
density distribution of the system is shallower in the central regions
than in the outer regions, as confirmed by the direct analysis of the
angle-averaged density profile of the object (see
Fig.~\ref{fig:prof_dens}).

\begin{figure}
  \centering
  \resizebox{1.00\hsize}{!}{\includegraphics[angle=-90]{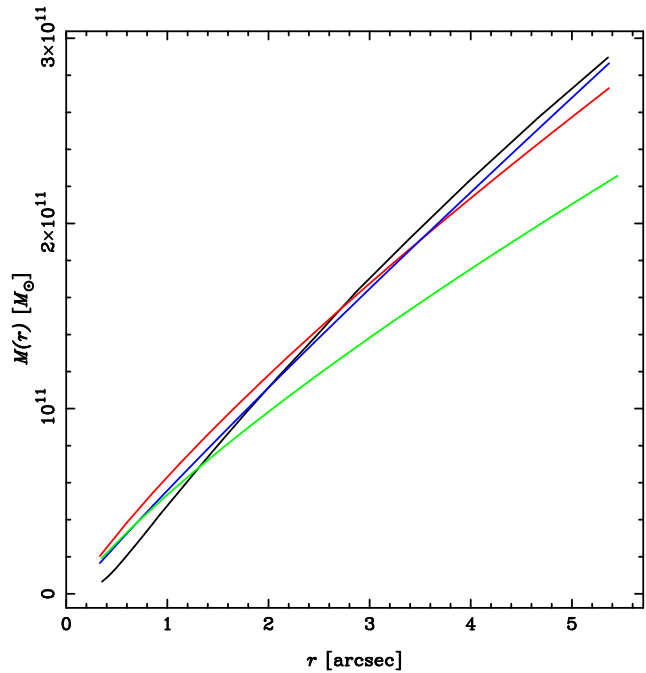}}
  \caption{Angle-averaged total mass distribution of the simulated
  system (black line) compared with the total mass profiles of the
  recovered best models (see Table~\ref{tab:eta}) for the three
  orthogonal projections data sets: {\yz} model (red line),
  {\zx} model (blue line) and {\xy} model (green line).}
  \label{fig:prof_mass}
\end{figure}

Despite several indications from the reconstructed observables that
the adopted power-law model is probably oversimplified for the data at
hand, we find that such a model still provides a satisfactory
description of the essential features of the system, and various
important physical quantities are robustly recovered.

The angle-averaged total density profiles corresponding to the best
models of Table~\ref{tab:eta} for the three data sets are plotted in
Fig.~\ref{fig:prof_dens} and compared with the true profile of the
simulated system. The density slope, which is very close to $\slope
\sim 2.2$, is quite accurately recovered beyond the inner $\sim 0.5$
arcsec. The {\zx} model presents a slightly shallower profile ($\slope
\sim 2.1$), although the discrepancy is small. The {\xy} model, which
in general provides the worst recovery of the true quantities (as it
will be further discussed in this Section), has a density
normalization which is lower than the real one, but manages to catch
almost perfectly the correct density slope. In the inner half
arcsecond, the density profile of the simulated galaxies becomes
shallower ($\rho \sim 1/r$), a feature that the power-law model is
obviously unable to capture, although, as already discussed, we have
other signals (e.g. from the velocity dispersion) that the model
breaks down in that approximate region, and that this is due to
the presence of a break in the density profile. Slightly more
sophisticated models, such as a double power-law or a single power-law
with an added break radius, have also been explored, but the
reconstruction of the observables does not improve significantly and
the additional complexity is therefore penalized by the Bayesian
evidence in favour of the single power-law. In light of what we know
from the direct examination of the simulated object, the reason for
this failure is that the transition from the $\rho \sim 1/r^2$ region
to the $\rho \sim 1/r$ region is too abrupt to be adequately
reproduced by these models, and therefore they do not perform
significantly better than the simpler single power-law.

Whereas the density slope is a substantially unharmed survivor of the
crash test, the recovered flattening and inclination angle are not
reliable parameters in case the systems deviate too drastically from
the model's assumptions. One should note, however, that these
quantities are only properly defined for an axisymmetric object, and
therefore do not have a straightforward interpretation when directly
applied to a simulated system which is approximately triaxial and
whose axis ratios also change as a function of radius (as seen in
Section~\ref{sec:simulation}). The best models for the {\yz} and {\xy}
data set both give an inclination close to $\sim 55\degr$, so there is
no sign that the latter is interpreted as a face-on system. The {\zx}
best model, on the other hand, turns out quite correctly to be an
almost edge-on system ($i \simeq 90\degr$). As for the flattening, it
appears that the axisymmetric model, faced with the insurmountable
problem of triaxiality, tends to adopt a density profile close to
spherical (almost exactly spherical in the case of the {\yz} data set,
slightly prolate or oblate in the cases of the $zx$- and {\xy} data
sets, respectively).


\begin{table}
  \centering
  \caption{Dark matter fraction within a sphere of radius $r = \Reff$
  (first column) and within the line-of-sight oriented cylinder of
  radius $R = \Reff$ (second column) for the three best models. The
  corresponding quantities for the true system within the same radius
  are also presented.}
  \smallskip
  \begin{tabular}{ c c c }
    \hline
    \noalign{\smallskip}
    & DM fraction & DM fraction \\
    & (sphere)    & (cylinder)  \\
    \noalign{\smallskip}
    \hline
    \noalign{\smallskip}
     {\yz} model ($\Reff = 5\farcs 23$) & 0.20 & 0.29 \\ 
     true system (at the same radius)        & 0.16 & 0.33 \\ 
    \noalign{\smallskip}
    \hline
    \noalign{\smallskip}
     {\zx} model ($\Reff = 5\farcs 36$) & 0.24 & 0.37 \\ 
     true system (at the same radius)        & 0.16 & 0.34 \\ 
    \noalign{\smallskip}
    \hline
    \noalign{\smallskip}
     {\xy} model ($\Reff = 6\farcs 16$) & 0.19 & 0.25 \\ 
     true system (at the same radius)        & 0.20 & 0.35 \\ 
    \noalign{\smallskip}
    \hline
  \end{tabular}
  \label{tab:DM}
\end{table}

\subsection{Total mass distribution and dark matter fraction}
\label{ssec:mass}

\begin{figure}
  \centering
  \resizebox{1.00\hsize}{!}{\includegraphics[angle=-90]{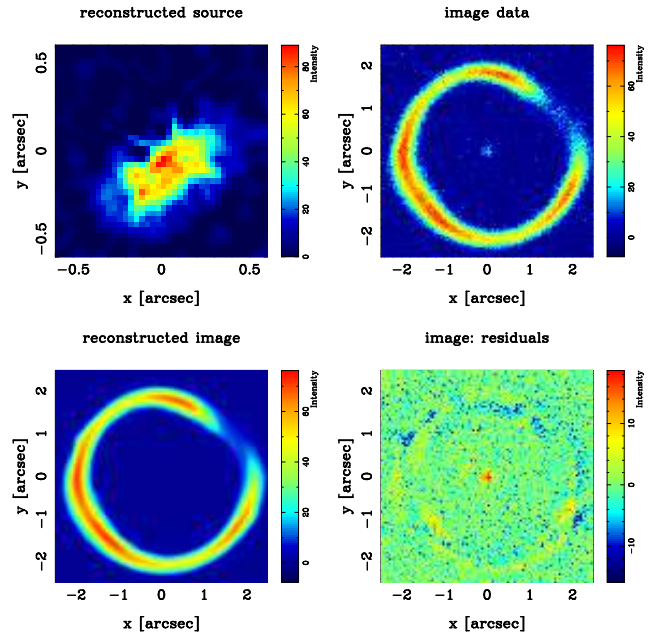}}
  \caption{Best model lens image reconstruction for the {\yz}
    data set (the generation of the simulated observables is detailed
    in Section~\ref{sec:simulation}). From the top left-hand to bottom
    right-hand panel: reconstructed source model; simulated noisy data
    showing the lensed image; lensed image reconstruction; residuals.}
  \label{fig:YZ-LEN}
\end{figure}
\begin{figure}
  \centering
  \resizebox{1.00\hsize}{!}{\includegraphics[angle=-90]{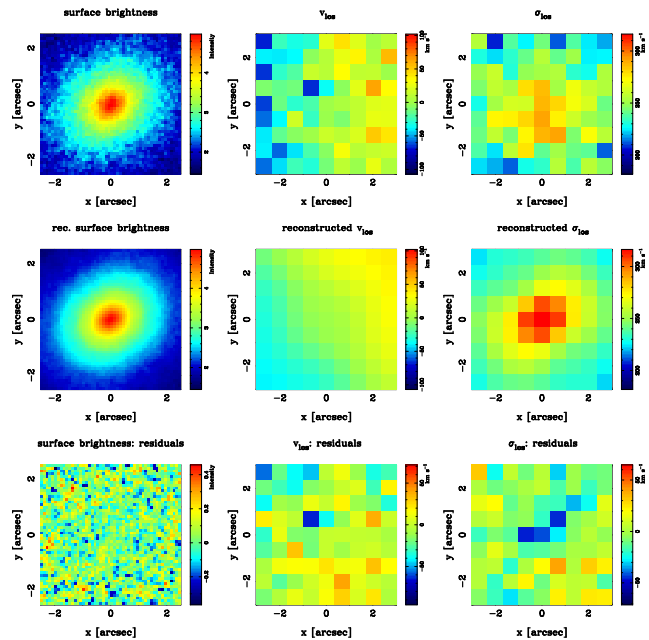}}
  \caption{Best dynamical model for the {\yz} data set. First
    row: simulated noisy surface brightness distribution, projected
    line-of-sight velocity and line-of-sight velocity
    dispersion. Second row: corresponding reconstructed quantities for
    the best model. Third row: residuals.}
  \label{fig:YZ-DYN}
\end{figure}

Closely connected to the density profile is the (angle-averaged) total
mass distribution, plotted in Fig.~\ref{fig:prof_mass} for the three
best models and the true system. We find that in the case of the
edge-on projections models ($yz$- and {\zx}s) the mass profile is very
well reproduced within a few percent. The mass profile of the face-on
projection model is instead underestimated of about $25$ percent
within a sphere of radius $\sim 5$ arcsec: this is a consequence of
the too low density normalization which is recovered for this model,
as previously seen in Fig.~\ref{fig:prof_dens}. Because of the tight
constraints imposed by the lensing, the total projected mass enclosed
within $\REin$ is within a few percent from the correct value for all
three models.

A quantity of extreme interest in the study of early-type galaxies is
the dark matter fraction of these objects. It is therefore important
to be able to test how reliable is the {\cauldron} method in
estimating this parameter also for systems that defy its assumptions
of axisymmetry and two-integral DF. Since the method only provides a
total density distribution, however, it is necessary to make further
assumptions to be able to constrain the stellar density profile. In
order to limit as much as possible the arbitrariness of such
assumptions we adopt, as in the analysis of real galaxies, the
so-called ``maximum bulge'' approach (cf. C08). This consists in
maximizing the contribution of the luminous component (which is
obtained as an output of the best model reconstruction),
i.e. maximally rescaling the stellar density distribution without
exceeding the total density distribution, assuming that the stellar
mass-to-light ratio is independent of position. This method gives a
lower limit for the dark matter fraction, provided that the model's
assumptions hold true.

\begin{figure}
  \centering
  \resizebox{1.00\hsize}{!}{\includegraphics[angle=-90]{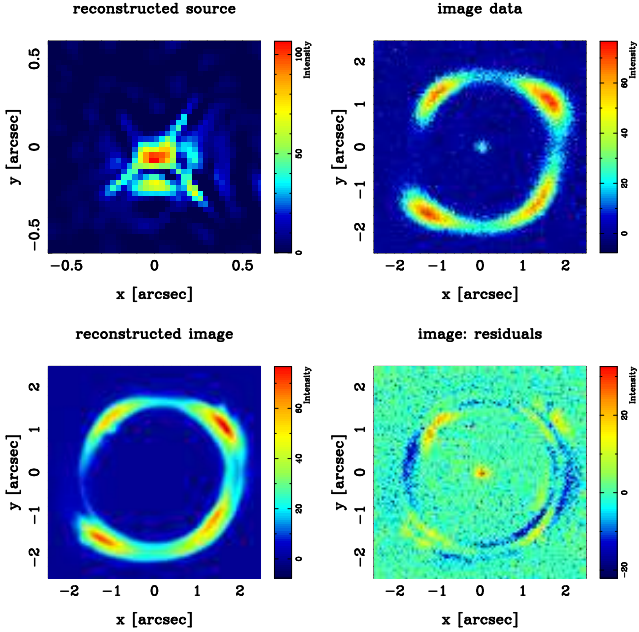}}
  \caption{Same as Fig.~\ref{fig:YZ-LEN}, but relative to the
  {\zx} data set.}
  \label{fig:ZX-LEN}
\end{figure}
\begin{figure}
  \centering
  \resizebox{1.00\hsize}{!}{\includegraphics[angle=-90]{fig6b.ps}}
  \caption{Same as Fig.~\ref{fig:YZ-DYN}, but relative to the
  {\zx} data set.}
  \label{fig:ZX-DYN}
\end{figure}

\begin{figure}
  \centering
  \resizebox{1.00\hsize}{!}{\includegraphics[angle=-90]{fig7a.ps}}
  \caption{Same as Fig.~\ref{fig:YZ-LEN}, but relative to the
  {\xy} data set.}
  \label{fig:XY-LEN}
\end{figure}
\begin{figure}
  \centering
  \resizebox{1.00\hsize}{!}{\includegraphics[angle=-90]{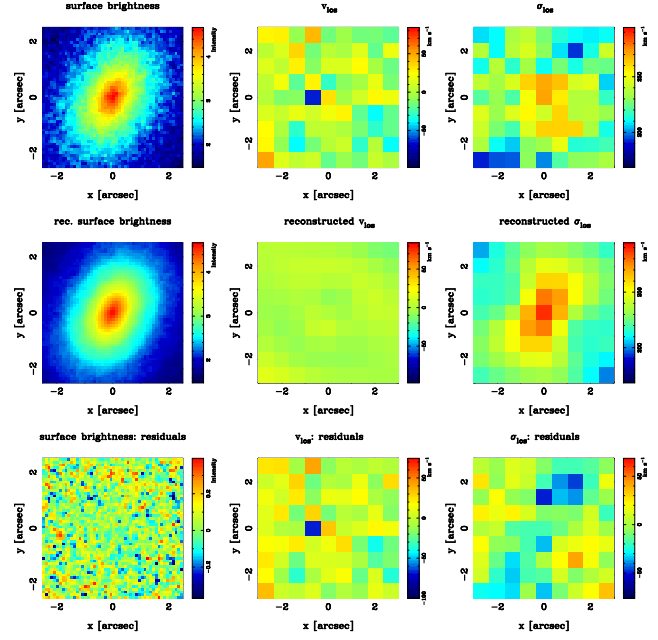}}
  \caption{Same as Fig.~\ref{fig:YZ-DYN}, but relative to the
  {\xy} data set.}
  \label{fig:XY-DYN}
\end{figure}

Under the maximum bulge hypothesis, we study both the volume mass
ratio (i.e. the dark matter fraction within a sphere of radius taken
to be equal to the effective radius of the considered data set) and
the projected mass ratio (i.e. the dark matter fraction within a
cylinder oriented along the line-of-sight with a radius equal to
$\Reff$). The results of this analysis for the three best models are
summarized in Table~\ref{tab:DM}, which also shows the corresponding
quantities for the true system (at the appropriate radii). We find
that the dark matter fraction is remarkably well recovered for all
three models, and it is within $10 \%$ (in total mass) of the correct
value for both the volume and the projected mass ratio. The largest
discrepancy ($10 \%$ within the cylinder) is found for the face-on
model, as usual the most problematic one, while the {\yz} model (which
is the one that best reproduces the observables, in particular the
lensed image) manages to accurately recover the dark matter fraction
within $\la 5 \%$ of the correct value.

It must be noted that we assumed a position-independent stellar
mass-to-light ratio both in constructing the surface-brightness map
from the N-body system and in estimating the maximum stellar mass for
the best-fit models. In real galaxies the stellar mass-to-light ratio
might depend on position, though the effect of a non-uniform stellar
mass-to-light ratio is not expected to be strong based on observed
colour gradients \citep[e.g.][]{Kronawitter2000}.


\subsection{Reconstructed source}
\label{ssec:source}

As it can be immediately seen by an examination of
Figs~\ref{fig:YZ-LEN}, \ref{fig:ZX-LEN} and~\ref{fig:XY-LEN}, the
lensed image reconstruction for the three models is far from the noise
level, and the reconstructed source -- with the partial exception of
the {\yz} model -- has little in common with the simulated source
(Fig.~\ref{fig:LENsource}) used to generate the lensed images. Even
assuming that one has no information about the actual source (as it
would be the case for real data sets) it is evident that the
reconstructed sources are unrealistically irregular and pixelized. We
remark that this is a consequence of having an underlying density
distribution for the system which is more complex and less homogeneous
than the assumed power-law density model.  As the present test neatly
displays, gravitational lensing is very sensitive to the features of
the potential within the Einstein radius, and even small
inhomogeneities and departures from the assumed model will generally
have a detectable effect\footnote{As a consequence, gravitational
lensing can actually be used to quantify the level of mass
substructure in massive galaxies, through their effect on
highly-magnified arcs and Einstein rings \citep[see e.g.][hereafter
VK08]{Vegetti2008}.}. Significantly, such large residuals have never
been encountered so far when analysing real lens galaxies, with all
the examined SLACS lenses being accurately reconstructed by means of
single power-law models, possibly with the inclusion of external shear
(\citealt{Koopmans2006}, \citealt{Czoske2008}, Barnab\`e et al., in
preparation). This strengthens the statement that the simulated system
under analysis, while not being utterly unrealistic, constitutes an
extreme case and is therefore well suited for the ``crash test'' of
the code that we aim to conduct. It should be emphasized that, if
these were real data sets, from the mere visual inspection of the
results of the best lensed image reconstruction provided by
{\cauldron}, we would already be able to conclude that we are dealing
with a unusually complex galaxy, which would surely deserve a more
sophisticated modelling once the preliminary study is completed.

\begin{figure*}
  \centering
  \resizebox{1.00\hsize}{!}{\includegraphics[angle=0]{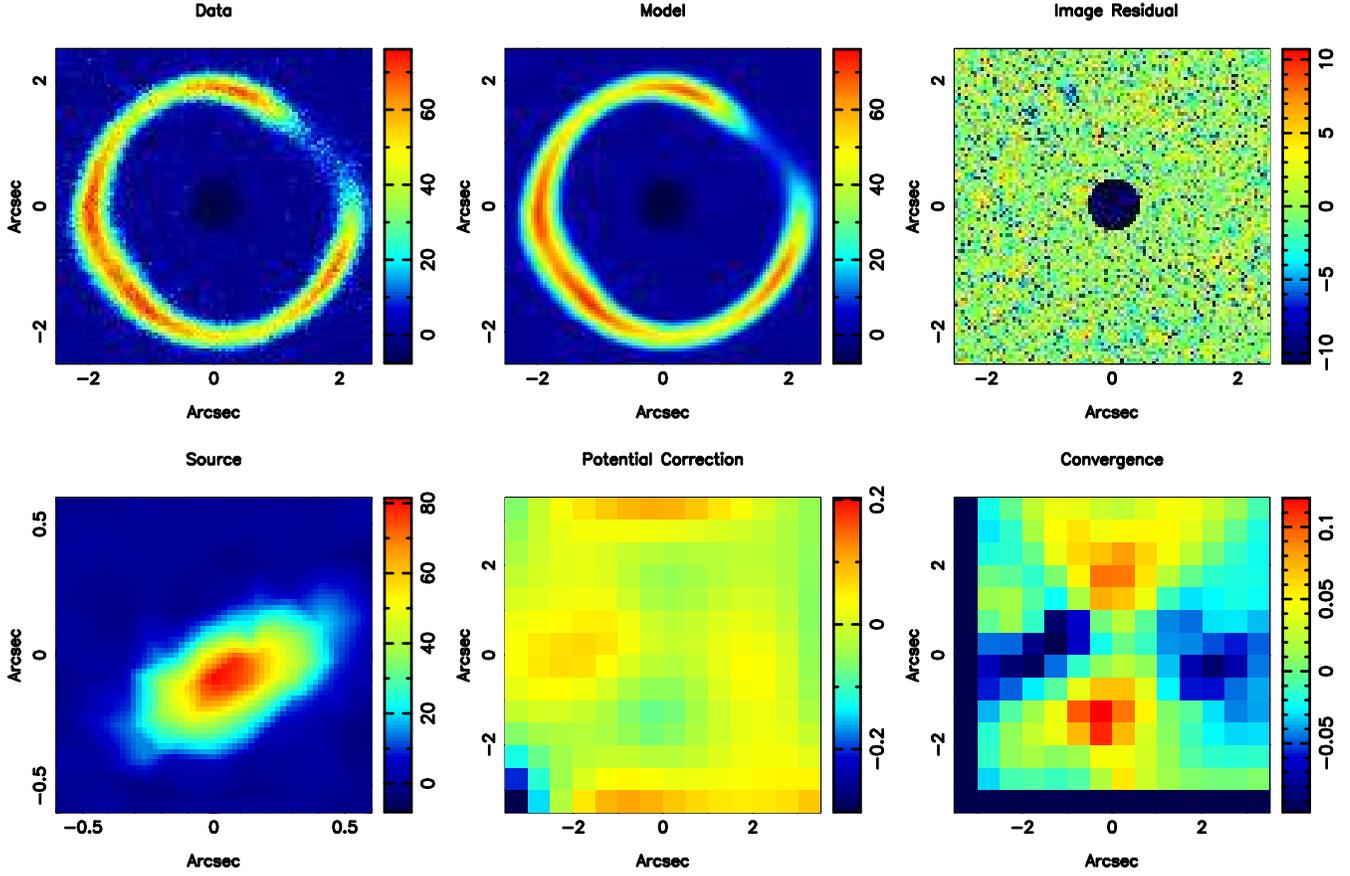}}
  \caption{Result of the linear source and potential reconstruction
  for the {\yz} dataset, as obtained by applying the adaptive
  Bayesian lensing code of VK08. The first row shows,
  from left to right, the lensed image data set, the reconstructed
  image and best source. The second row presents, from left to right,
  the image residuals, the total potential correction and the
  substructure convergence.}
  \label{fig:adaptive}
\end{figure*}

In order to better understand the limitations of the axisymmetric
approach and the cause for the large residuals in the
reconstructed image, the same data set (i.e. {\yz} projection) has
been analysed by means of the adaptive Bayesian strong lensing code of
VK08, which can account for small corrections in the projected
potential, and therefore for departures from symmetry and for
the presence of clumpiness and substructure in the convergence, but
which does not include any constraint from the dynamics. Starting from
the best model axisymmetric potential of Table~\ref{tab:eta} and then
introducing small potential corrections and letting the parameters
vary, it is found that the lensed image can be reconstructed down to
the noise level, and the source is quite accurately recovered, by
making use of an adaptive grid, as shown in
Fig.~\ref{fig:adaptive}. As it is visible in the convergence map of
Fig.~\ref{fig:adaptive} (bottom-right panel), there are two main
overdensities (with respect to the best model density profile),
located on opposite sides with respect to the center. These
overdensities are at least an order of magnitude too weak to be
indicative of the presence of a genuine localized massive
dishomogeneity in the simulated system (cf. VK08 and in particular
their Figs.~8 and~9), which is rather smooth and does not present
massive substructures. These features in the convergence map are
therefore due, rather than to clumpiness, to the deviation of the true
(projected) mass distribution from the simple assumption of elliptical
shape.

\subsection{Recovered dynamical quantities}
\label{ssec:dynamics}

A reliable knowledge of the dynamical structure of early-type galaxies
would constitute a valuable asset for all the formation and evolution
models. Therefore, in this Section we investigate how accurately the
axisymmetric {\cauldron} code can recover the essential dynamical
characteristics of the simulated system. We expect comparable
performances (and typically better ones) when the code is applied to
real galaxies.

Since the dynamical modelling block of {\cauldron} (see
Section~\ref{sec:code}) does not employ an actual orbit-superposition
method, a detailed analysis of the orbital families of the galaxy is
beyond reach. However, we are able to study fundamental global
dynamical quantities such as angular momentum, $V/\sigma$ and
anisotropy parameters. The results of this study for the three models,
compared with the true quantities calculated directly from the
simulated system, are listed in Table~\ref{tab:dyn}.

The total angular momentum along the principal rotation axis, $\Lz$,
has been calculated for the three models and the simulated system
within the same cylindrical region of radius $R = 5 \arcsec $ and
height $z = 5 \arcsec$ (this is a square box in the meridional plane,
of linear size comparable to the effective radius; all the dynamical
quantities have been calculated within this region). As seen in
Table~\ref{tab:dyn}, $\Lz$ is accurately recovered for the two edge-on
projection models, with discrepancies of less than $10 \%$ from the
correct value, despite the fact that the line-of-sight velocity maps
are considerably noisy\footnote{The mock galaxy displays also some
rotation along the orthogonal axes, although the angular momentum
along these directions is only of order \mbox{$\la 1 \%$} of
$\Lz$. Because of its intrinsic properties, the model is not capable
of reproducing this kind of rotation (in other words, $L_{x}$ and
$L_{y}$ are always $0$ by construction).}. On the contrary, in the
case of the face-on projection model, where little or no information
about the rotation is available from the kinematic data (see
top-middle panel of Fig.~\ref{fig:XY-DYN}: the map displays
essentially no rotation) the recovered $\Lz$ is obviously incorrect,
being of the opposite sign and, more importantly, quite close to
zero. This clearly highlights the great importance of the information
enclosed in the kinematic maps, despite the fact that such maps are
usually much more noisy and considerably more coarsely sampled, in
terms of number of pixels, than the surface brightness or lensed image
maps. On the opposite side, this result also cautions us in being
aware of the possible shortfalls of the method when studying galaxies
whose data show no discernible rotation.

\begin{table}
  \centering
  \caption{Recovered dynamical quantities for the three best models
    (last three columns) compared with the true values directly
    calculated from the N-body system (second column). The dynamical
    quantities are $\Lz$ (in units of kpc km s$^{-1}$), the $V/\sigma$
    ratio, and the three global anisotropy parameters $\delta$,
    $\beta$ and $\gamma$. See text for a more exhaustive description.}
  \smallskip
  \begin{tabular}{ c c c c c }
    \hline
    \noalign{\smallskip}
     & true value & {\yz} & {\zx} & {\xy}       \\
    \noalign{\smallskip}
    \hline
    \noalign{\smallskip}
    $\Lz$      & -143.1 & -142.6     & -143.3     &   16.8     \\
    $V/\sigma$ &  0.170 &  0.152     &  0.214     &  0.130     \\
    $\beta$    &  0.301 & $\equiv 0$ & $\equiv 0$ & $\equiv 0$ \\
    $\gamma$   &  0.208 & -0.470     & -0.645     & -0.986     \\
    $\delta$   &  0.219 &  0.190     &  0.244     &  0.330     \\
    \noalign{\smallskip}
    \hline
  \end{tabular}
  \label{tab:dyn}
\end{table}

The global quantity $V/\sigma$ \citep[see][]{Binney2005,SauronX} is an
indicator of the importance of rotation with respect to the random
motion. We find here that a value not too far from the correct one is
recovered for the edge-on projection models, although the
discrepancies are larger than in the case of the angular momentum (the
difference is of order $10 \%$ for the {\yz} model and $25 \%$ for the
{\zx} model), and even the face-on projection model, while
underestimating the importance of rotation of about $25 \%$, gives a
quite reasonable result.

The anisotropy distribution is another very relevant dynamical
quantity, often considered an important indicator of the galaxy
formation processes. Since the simulated system is severely
non-spherical, however, the anisotropy profile cannot be reliably
described and compared to the models by making use of a simple radial
parameter such as the commonly used $\beta_{r}(r) \equiv 1 -
\sigma^{2}_{\rm tan}(r) / \sigma^{2}_{\rm rad}(r)$, where $\sigma_{\rm
tan}$ and $\sigma_{\rm rad}$ are the tangential and radial velocity
dispersioni, respectively. Instead, a more robust and consistent
indicator is provided by the three global anisotropy parameters
defined in \citet{SauronX} and \citet{BT08}:
\begin{equation}
  \label{eq:AP:beta}
  \beta \equiv 1 - \frac{\Pi_{zz}}{\Pi_{RR}}, 
\end{equation}
\begin{equation}
  \label{eq:AP:gamma}
  \gamma \equiv 1 - \frac{\Pi_{\varphi\varphi}}{\Pi_{RR}}, 
\end{equation}
\begin{equation}
  \label{eq:AP:delta}
  \delta \equiv 1 - \frac{2 \Pi_{zz}}{\Pi_{RR} + \Pi_{\varphi\varphi}} = 
  \frac{2 \beta - \gamma}{2 - \gamma},
\end{equation}
where 
\begin{equation}
  \label{eq:AP:PI}
  \Pi_{kk} = \int \rho \sigma^{2}_{k}\, \mathrm{d}^{3}x 
\end{equation}
and $\sigma_{k}$ denotes the velocity dispersion along the direction
$k$ at any given location in the galaxy. For a two-integral DF
$\sigma_{R}^{2} = \sigma_{z}^{2}$ everywhere, which implies $\Pi_{RR}
= \Pi_{zz}$, so that the value of $\beta$ is always zero. Therefore,
due to the simplifying assumptions on the distribution function, our
method will generally fail in recovering the true $\beta$ value when
analysing a more complex system. We clearly observe this in the
present study, where $\beta = 0.301$ for the simulated system (see
again Table~\ref{tab:dyn}). Not too unexpectedly, since the global
anisotropy parameters are related, this has disrupting consequences
also on the recovered $\gamma$, which is always negative,
indicating mild tangential anisotropy in the models, while the mildly
radially anisotropic N-body system has a positive $\gamma$. This
discrepancy is confirmed by inspecting the radial behaviour of a
kinetic energy proxy for $\beta_{r}(r)$, i.e. the quantity
$\beta_{K}(r) \equiv 1 - K_{\rm tan}(r) / K_{\rm rad}(r)$, where $
K_{\rm tan}$ and $K_{\rm rad}$ are, respectively, the
spherically-averaged tangential and radial components of the kinetic
energy.  Interestingly, however, the global parameter $\delta$ is
remarkably robust, particularly for the two edge-on projection models,
with a discrepancy of order $\sim 15 \%$ from the true value.


\begin{figure*}
  \centering
  \resizebox{1.00\hsize}{!}{\includegraphics[angle=-90]{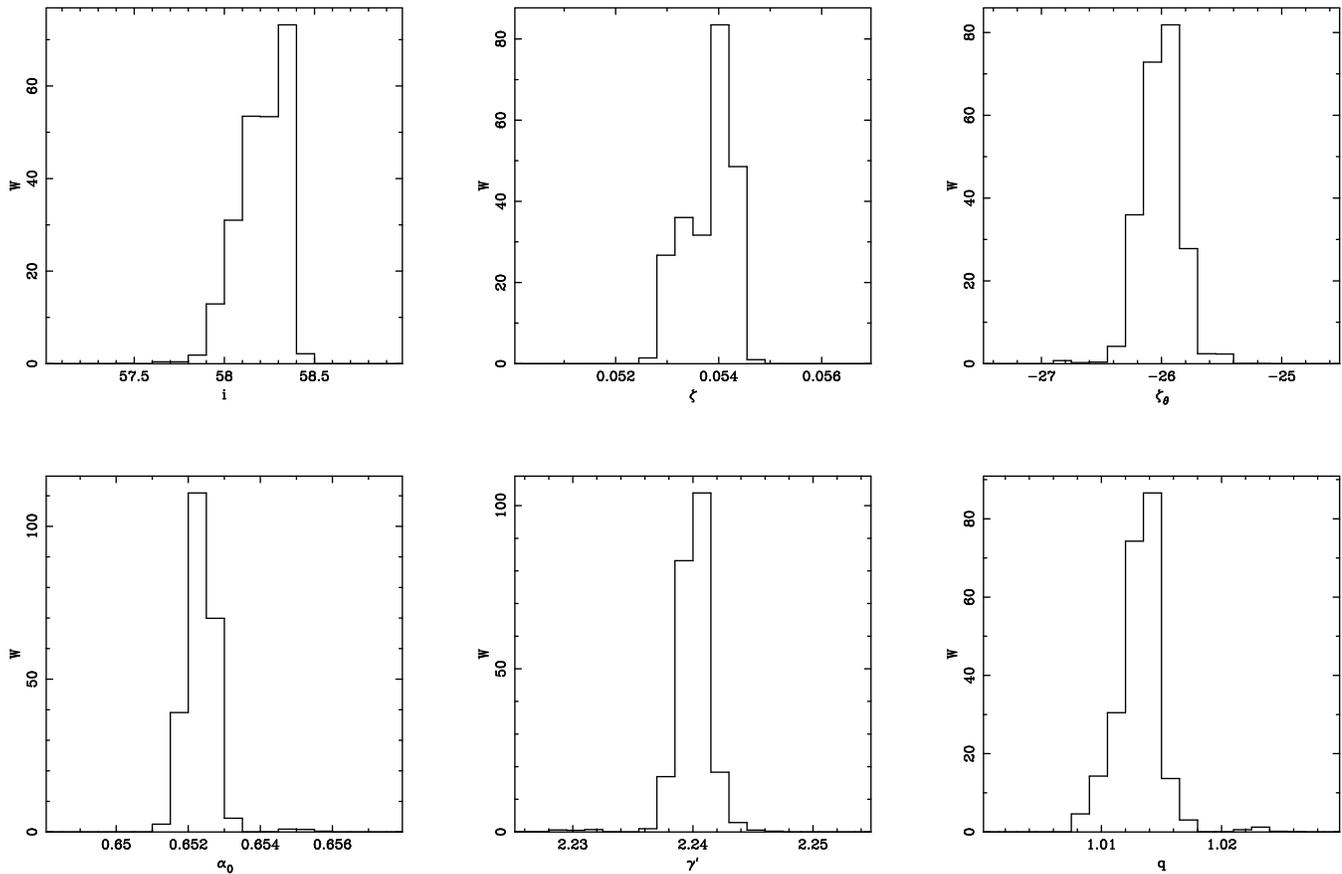}}
  \caption{Posterior probability distribution for the non-linear
  parameters of the best power-law model for the {\yz} projection data
  set, as obtained from the Nested Sampling evidence
  exploration. The width of the posterior probability
  distribution has been calculated for each of the parameters, by
  considering the region around the peak which contains 99\% of the
  probability, yielding: $i = [57.81, 58.45]$, $\shear = [0.05273,
  0.05455]$, $\angshear = [-26.55, -25.47]$, $\talp = [0.6512,
  0.6550]$, $\slope = [2.231, 2.244]$, $q = [1.008, 1.023]$.}
  \label{fig:YZ-NSerr}
\end{figure*}

\subsection{Uncertainties}
\label{ssec:errors}

In this section we present the statistical uncertainties on the model
parameters calculated by making use of the identical procedure which
is used when dealing with real systems.

Bayesian statistics represents a powerful tool for data analysis,
model comparison and model parameters constraining. A major
improvement in this direction has been made with the introduction of
the Nested Sampling technique developed by \citet{Skilling2004} (see
also \citealt{SS2006} and e.g. \citealt*{Mukherjee2006} for an
astrophysical application). This method provides a computationally
efficient way to calculate the total marginalised evidence, which is
the key quantity for model comparison, and in addition yields other
valuable information, such as the posterior probability density
distributions, the mean values and the standard deviations for each of
the model parameters. The Nested Sampling technique has been applied
in the context of lensing and model-comparison by VK08.

Bayesian statistics requires to formalize one's assumptions by
defining priors $P$ on the parameters $\eta_{\mathrm{i}}$ \citep[see
e.g.][]{MacKay1992}. We choose the priors to be uniform in a symmetric
interval of size $\delta \eta_{\mathrm{i}}$ around the best recovered
values $\eta_{\mathrm{b,i}}$, that is:
  \begin{equation}
    P\left(\eta_i\right)= \left\{
    \begin{array}{ll}
      \mathrm{constant} & \mathrm{for} \quad 
                          | \eta_{\mathrm{b,i}} - \eta_{\mathrm{i}} |
			  \leq \delta \eta_{\mathrm{i}} \\ 
        & \\ 
      0 & \mathrm{for} \quad | \eta_{\mathrm{b,i}} - \eta_{\mathrm{i}} |
          > \delta \eta_{\mathrm{i}} .
    \end{array} 
    \right.
  \end{equation}
In order to make sure that the priors include the bulk of the evidence
likelihood, very conservative estimates of the intervals $\delta
\eta_{\mathrm{i}}$ are obtained by means of fast preliminary runs.

The posterior probability distributions (PPDs) for each parameter
obtained from the Nested Sampling analysis are shown in
Fig.~\ref{fig:YZ-NSerr} for one of the models ({\yz}
projection). Within the context of Bayesian statistics, each PPD
histogram quantifies the error for the considered parameter given the
data and all the assumptions (i.e. under the hypothesis that the
adopted model is the correct one); since the many pixels in the data
sets provide a lot of constraints, these errors are typically very
small, as is the case in the plot presented here. It should be noted
that, because of the marginalization over all the parameters except
the one under analysis, the PPD usually provides the most conservative
estimate of statistical errors. At the same time, however, due to
projection effect arising from the marginalization itself, the mean
value of the parameter obtained from the PPD can be significantly
skewed with respect to the best recovered value of the corresponding
parameter obtained from the best model (P. Marshall, private
communication). 

However, the statistical errors cannot take into account or give an
estimate of the systematic uncertainties, which are frequently much
larger than the former. In our case, significant systematic errors
arise mostly due to the adoption of an oversimplified
model\footnote{In real datasets of lens galaxies, sources of
systematic errors which raise particular concern are, for example, the
PSF and the subtraction of the lens galaxy surface brightness
\citep[see e.g. the detailed study of][]{Marshall2007}.}. The entity
of the systematic uncertainties can be quantified, at least as a first
order approximation, by looking at the discrepancies between the
recovered parameters for the three data sets best models.


\section{Discussion and conclusions}
\label{sec:conclusions}

We have applied {\cauldron}, currently the most advanced code for
joint gravitational lensing and stellar dynamics analysis of
early-type galaxies, to a galaxy model with dark matter halo resulting
from a numerical N-body simulation of galaxy merging. Such a N-body
system, which we use as lens, significantly violates the two major
assumptions upon which the algorithm is based, namely axial symmetry
of the total density distribution and two-integral stellar
distribution function. The purpose of this crash test is to
investigate how the code will perform or fail in an extreme case, and
to identify the quantities which can still be reliably recovered. Such
robust quantities are expected to be recovered with at least
comparable accuracy when {\cauldron} is employed in the analysis of
real galaxies which, while not necessarily axisymmetric or described
by a two- (or three-) integral DF, will hardly depart from the code's
assumptions more severely than the simulated system studied here.

Further complications can also arise, in general, due to the
effects of the environment on the lens galaxy. However, the SLACS
galaxies, despite living in overdense regions as expected for massive
early-type galaxies, are not found to be significantly affected by the
contribution of the environment or of line-of-sight contaminants
\citep[see the in-depth study of][]{Treu2008}. Therefore, since the
environment does not play a major role for lens galaxies at least in
the redshift range $z \sim 0.1 - 0.4$, we have not considered this
issue further in the present paper.

From the N-body system we have generated three data sets corresponding
to three orthogonal lines of sight, one of which has been chosen to be
approximately oriented along the total angular momentum of the
system, in order to obtain a ``face-on'' projection data set with
little or no rotation discernible in the kinematic maps. An elliptical
Gaussian surface brightness distribution has been constructed and then
gravitationally lensed by the mock galaxy in order to create the
lensing data set. We have also taken into account the effect of the
PSF and added realistic noise to the simulated data, using as a
reference the real data set for the SLACS lens galaxy {\galaxy}
studied in C08.

These data sets have been analysed with {\cauldron}, assuming as a
model an axisymmetric power-law total density distribution, with the
identical procedure followed in the study of real lens galaxies. In
the three cases, the recovered best models, obtained via maximization
of the Bayesian evidence, show clear difficulties in reconstructing
the observables (in particular the lensed image and the velocity
dispersion map) up to the noise level. This is a consequence of having
adopted a very simple model which cannot account for the complexity
and the lack of symmetry of the true density distribution.

However, the method is still capable of recovering with remarkable
accuracy several global structural and dynamical characteristics of
the examined system, provided that some information about the galaxy
rotation is available from the kinematic maps. In particular:

\begin{enumerate}
\item The logarithmic slope $\slope$ of the total density distribution
is a robust quantity which is recovered with remarkable accuracy
(within less than $10 \%$), even in the case of the face-on projection
data set. While a power-law model cannot account for a break in the
actual density profile, indications of its presence will show up in
the observables as features which are not reproduced by the best model
(e.g. a much flatter central velocity dispersion).
\item The angle-averaged total density and total mass radial profiles
for the edge-on projection best models are found to closely follow the
corresponding true distributions, within approximately an effective
radius. Discrepancies are larger for the face-on projection case.
\item The best reconstructed inclination angle and flattening of the
total density distribution have little or no relation with the
corresponding quantities of the N-body system. We conclude that when
the axisymmetric model is applied to a system with a more complex
geometry, one can not expect to recover reliable information about its
shape.
\item By adopting the maximum bulge approach (i.e. maximizing the
contribution of the luminous component, assuming a
position-independent stellar mass-to-light ratio), it is possible to
estimate within approximately $10 \%$ (in total mass) the dark matter
fraction of the analysed system, whether the mass ratio is calculated
within a sphere or within a line-of-sight oriented cylinder of radius
$\simeq \Reff$.
\item When rotation is present in the kinematic maps, global
quantities such as the angular momentum $\Lz$ and the ratio $V/\sigma$
(a measure of ordered vs chaotic motions) recovered from the best
model describe quite reliably (i.e. within $\sim 10 \%$ and $25 \%$,
respectively) the dynamical properties of the system under
analysis. The global anisotropy parameters $\beta$ and
$\gamma$ are not correctly estimated, i.e. the anisotropy
distribution is not robustly recovered by the method unless the
analyzed system effectively respects the assumptions of axisymmetry
and two-integral distribution function. The anisotropy
parameter~$\delta$, on the contrary, is found to be a robust quantity
even when such assumptions are violated: we recover the correct value
within $\la 15 \%$ for the edge-on projection data sets.
\end{enumerate}

The major conclusion of our study is that the joint lensing and
dynamics code {\cauldron} can be effectively applied also to the
analysis of galaxies which deviate from the (quite restrictive)
assumptions of axial symmetry and two-integral stellar DF. This result
is very relevant for the analysis of galaxies at $z \gtrsim 0.1$ for
which, due to the data limitations, the more powerful methods available
at lower redshifts are not viable techniques.  Several fundamental
structural and dynamical quantities, in particular the total density
slope and the dark matter fraction within the region where the data
are available, can be recovered with good accuracy. Other quantities,
such as $\Lz$ and the anisotropy parameter $\delta$, can be reliably
recovered provided that rotation is detected in the kinematic maps:
when this is not the case (either because of intrinsic properties of
the galaxy or because the system is observed face-on) the constraints
are much looser and one needs to be very sceptical about the outcomes
of the best model relative to these quantities. We point out that
special care should be taken when the best model does not manage to
reproduce the observables satisfactorily (i.e. at or close to the
noise level), in particular the lensed image. This is a strong
indication that the true density distribution deviates significantly
from the assumptions of the adopted model family, and therefore, while
there are still reliable recovered parameters (as listed above), more
delicate quantities such as flattening and inclination angle, as well
as the reconstructed source, should not be trusted. In these cases,
the {\cauldron} code can provide a robust but only quite general
description of the structure of the galaxy, paving the way for the
more sophisticated modelling techniques which are necessary (together,
if possible, with a better data set) to achieve a deeper and more
detailed knowledge of the system.

Interestingly, large residuals in the lensed images reconstruction
such as the ones found in this study have never been encountered so
far in the analysis of the SLACS sample of lens galaxies
(\citealt{Koopmans2006}, C08) indicating that an axisymmetric single
power-law total density distribution constitutes a satisfactory model
for these systems.


\section*{Acknowledgments}

M.~B. acknowledges the support from an NWO program subsidy (project
number 614.000.417). L.~V.~E.~K. and S.~V. are supported (in part)
through an NWO-VIDI program subsidy (project number
639.042.505). M.~B. and S.~V. are grateful to Phil Marshall for useful
discussion.

\bibliography{ms}

\label{lastpage}

\clearpage

\end{document}